%% file: main.tex
\documentclass[a4paper,11pt]{article}
\usepackage{jinstpub} 
\usepackage{lineno}
\usepackage{xcolor}

\title{\boldmath Hit-rate capability of a silicon strip detector module for decay positron detection in
the J-PARC muon $g-2$/EDM experiment}

\author[a]{Ryuto Azuma}
\author[b]{Katsunori Awa}
\author[b]{Shunsuke Doi}
\author[c]{Yowichi Fujita}
\author[d]{Seiso Fukumura}
\author[e]{Yu Goto}
\author[c]{Ryotaro Honda}
\author[f]{Sohtaro Kanda}
\author[g]{Tetsuichi Kishishita}
\author[h]{Tatsuya Kume}
\author[c,i,j]{Tsutomu Mibe}
\author[a]{Yukiharu Murata}
\author[f]{Shoichiro Nishimura}
\author[c]{Shinji Ogawa}
\author[c]{Yuta Okazaki}
\author[c,j]{Naohito Saito}
\author[b]{Maki Sakakibara}
\author[c]{Osamu Sasaki}
\author[j]{Taiki Sato}
\author[a,d]{Yutaro Sato}
\author[k]{Yoshiaki Seino}
\author[c]{Hiroshi Sendai}
\author[f]{Koichiro Shimomura}
\author[b]{Shohei Shirabe}
\author[c]{Masayoshi Shoji}
\author[f]{Patrick Strasser}
\author[l]{Taikan Suehara}
\author[e]{Shiori Sugahara}
\author[c]{Junichi Suzuki}
\author[h]{Toshikazu Takatomi}
\author[c]{Manobu M. Tanaka}
\author[b,m]{Junji Tojo}
\author[n]{Hiroyuki A. Torii}
\author[o,m,1]{Takashi Yamanaka\note{Corresponding author.}}
\author[c]{Hiroshi Yamaoka}
\author[f]{Takayuki Yamazaki}
\author[m,o]{Tamaki Yoshioka}

\affiliation[a]{Department of Fundamental Sciences, Graduate School of Science and Technology, Niigata University,
    950-2181,
    Niigata,
    Niigata,
    Japan}
\affiliation[b]{Department of Physics, Graduate School of Science, Kyushu University,Nishi-ku,
    Fukuoka,
    819-0395,
    Japan}
\affiliation[c]{Institute of Particle and Nuclear Studies, High Energy Accelerator Research Organization,
    Tsukuba,
    Ibaraki,
    305-0801,
    Japan}
\affiliation[d]{Department of Physics, Faculty of Science, Niigata University,
    Niigata,
    Niigata,
    950-2181,
    Japan}
\affiliation[e]{Graduate School of Science, Nagoya University,
    Chikusa-ku,
    Nagoya,
    464-8602,
    Japan}
\affiliation[f]{Institute of Materials Structure Science, High Energy Accelerator Research Organization, Tsukuba, Ibaraki 305-0801, Japan}
\affiliation[g]{Physikalisches Institut, University of Bonn, Bonn,  53115,  Germany}
\affiliation[h]{Mechanical Engineering Center, High Energy Accelerator Research Organization,
    Tsukuba,
    Ibaraki,
    305-0801,
    Japan}
\affiliation[i]{International Center for Quantum-field Measurement Systems for Studies of the Universe and Particles, High Energy Accelerator Research Organization, Tsukuba, Ibaraki, 305-0801, Japan}
\affiliation[j]{Department of Physics, Graduate School of Science, The University of Tokyo,
    Bunkyo-ku,
    Tokyo,
    113-0033,
    Japan}
\affiliation[k]{National Institute of Technology, Toyama College,
    Imizu,
    Toyama,
    933-0293,
    Japan}
\affiliation[l]{International Center for Elementary Particle Physics, The University of Tokyo,
    Bunkyo-ku,
    Tokyo,
    113-0033,
    Japan}
\affiliation[m]{Research Center for Advanced Particle Physics, Kyushu University,
    Nishi-ku,
    Fukuoka,
    819-0395,
    Japan}
\affiliation[n]{School of Science, The University of Tokyo,
    Bunkyo-ku,
    Tokyo,
    113-0033,
    Japan}
\affiliation[o]{Faculty of Arts and Science, Kyushu University,
    Nishi-ku,
    Fukuoka,
    819-0395,
    Japan}
    
\emailAdd{yamanaka@artsci.kyushu-u.ac.jp}

\abstract{In the J-PARC muon $g-2$/EDM experiment, a silicon strip detector
  will be used to detect positrons from muon decays. The detector
  consists of planes of detector modules arranged radially.
  The expected maximum hit rate reaches 1.4~MHz per sensor strip,
  and achieving high detection efficiency even under such hit-rate conditions is a key performance requirement.
  We have developed the smallest unit of the detector module,
  and its performance was evaluated using a muon beam at 
  the J-PARC MLF H-line.
  The specifications of the detector module and the evaluated 
  hit-rate capability are described in this article.
  }

\keywords{Si microstrip and pad detectors, Particle tracking detectors (Solid-state detectors)}

\arxivnumber{2606.16347} 

\begin{document}
\maketitle
\flushbottom

\input{Introduction}

\input{Detector}

\input{Experiment}

\input{Performance}

\section{Conclusions}
\label{sec:conclusions}
A silicon strip detector module was developed for decay positron
tracking at the J-PARC muon $g-2$/EDM experiment.
The expected maximum hit rate reaches 1.4~MHz per sensor strip
in this experiment, and achieving high detection efficiency even 
under such hit-rate conditions is a key performance requirement.
The hit-rate capability of the developed detector module
was evaluated using the MuSEUM experimental setup at the J-PARC MLF H-line.
The loss of signal hits due to pileup was observed almost proportional to
the hit rate, and the efficiency loss at the hit rate of 1.4~MHz
was measured to be 10\%. Considering the expected number of hits
from positron tracks in the muon $g-2$/EDM experiment, a 10\% loss of hits
under the highest hit-rate conditions has a negligible impact on the track reconstruction
efficiency, indicating that the developed
detector module satisfies the performance requirements of the experiment.

\acknowledgments
The authors would like to thank the KEK and J-PARC muon section staff for their
strong support, and the Open Source Consortium of Instrumentation (Open-It) of KEK 
for their support on the electronics design. This work is supported by JSPS KAKENHI 
Grants No. JP15H05742, No. JP20H05625, No. JP23K22503, No. JP25H01292 and 
No. JP26K21728.


\bibliographystyle{elsarticle-num}
\bibliography{references}




\end{document}

%% file: Introduction.tex
\section{Introduction}
\subsection{Muon $g-2$ and EDM}
Precision measurements of the muon anomalous magnetic moment ($g-2$) and
the electric dipole moment (EDM) are important probes for searching for 
physics beyond the Standard Model (SM). The recent measurement by the FNAL E989 experiment
determines the muon $g-2$ with a precision of 127~ppb~\cite{FNAL_E989}, and
this result is consistent with the previous measurement reported by the BNL E821 
experiment~\cite{BNL_E821}.
With this result, the tension between the experimental average and the 
SM prediction based on $e^{+}e^{-}$ experimental data
for hadronic vacuum polarization exceeds 5~$\sigma$. However, the SM prediction based on lattice QCD calculations
is found to be consistent with the experimental average, 
with Ref.~\cite{g-2_white_paper} cited as a representative result.
The muon EDM was searched for by the BNL E821 experiment, and the current upper limit 
has been set to
$|d_{\mu}|<1.8\times10^{-19}$~$e\cdot$cm~\cite{BNL_EDM}. A new experiment employing the so-called frozen-spin technique
has been proposed at PSI, aiming for a sensitivity of $6\times 10^{-23}$~$e\cdot$cm~\cite{PSI_EDM}.

\subsection{Muon $g-2$/EDM experiment at J-PARC}
Under these circumstances, a new experiment has been proposed at J-PARC to measure the muon $g-2$ and EDM
using a technique different from those employed at the BNL and FNAL experiments~\cite{J-PARC_E34}.
The key feature of this experiment is the acceleration of a thermal muon beam.
After producing a high-intensity surface muon beam, it is cooled to room temperature by
forming muounium in an aerogel target, and then re-accelerated to a momentum of 300~MeV/$c$.
By injecting this muon beam into a 3~T storage magnet, strong focusing by an electric field is not required. Instead, a weak focusing
magnetic field is sufficient to store the beam on a stable orbit.
This technique enables the removal of the electric fields in the storage volume,
which otherwise cause systematic biases in the $g-2$ measurement. Furthermore, eliminating the
electric field removes the requirement to operate at the so-called magic momentum
of approximately 3.09~GeV/$c$ in order to cancel the electric field contribution to the $g-2$ measurement.
By using a relatively low momentum muon beam of 300~MeV/$c$,
it becomes possible to employ a highly uniform magnetic field and a compact storage ring with a diameter
of 66~cm. This compact storage magnet also enables the use of
a full tracking detector to detect positrons from muon decays.

\subsection{Requirements and design of the tracking detector}
The experiment requires a decay-positron detector capable of detecting positrons
following helical orbits in a 3~T magnetic field within a cylindrical volume of  
290~mm radius and 400~mm height. Positrons with momenta
above 200~MeV/$c$, which are relevant to the $g-2$ measurement,
must be reconstructed at a
 maximum instantaneous rate corresponding to six muon decays per nanosecond, 
while accommodating the time-dependent decay rate associated
with the muon lifetime
at 300~MeV/$c$ ($\sim 6.6~\upmu$s).
The detector must tolerate this hit rate and
measure signals up to four minimum ionizing particle (MIP)
charges to accommodate hit pileup. A signal-to-noise
ratio greater than 15 is required to achieve high-efficiency
positron-track reconstruction.
The time walk must be less than 1~ns over a signal range 
corresponding to 0.5--3~MIP charges to minimize
shifts in the recorded hit times caused by pileup.

\begin{figure}[htbp]
\centering
\includegraphics[width=0.8\textwidth]{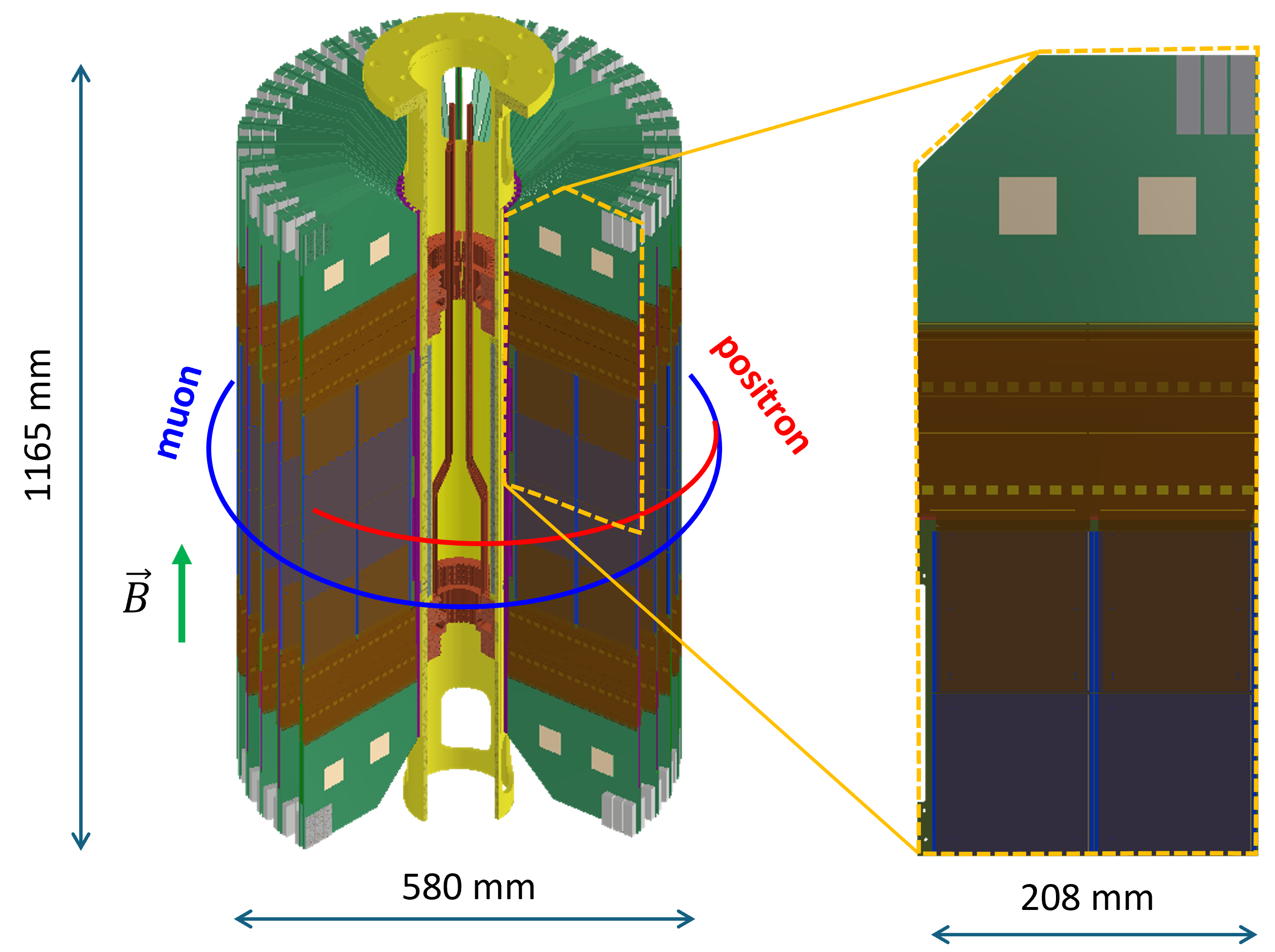}
\caption{Cut-out view of the positron tracking detector (left) and the smallest unit of
the detector module; quarter vane (right).}\label{fig:positron_detector}
\end{figure}

To meet these requirements, a tracking detector composed
of silicon strip detector modules
was designed, as shown in Figure~\ref{fig:positron_detector}.
The detector consists of 40 vane-like modules arranged radially.
The smallest unit of the detector corresponds to one quarter of 
the vane-like module and is therefore referred to as a ``quarter vane''.
We have developed a quarter vane module using silicon strip sensors with a
strip pitch of 190~$\upmu$m and a length of 48.365~mm.
With this design, the maximum hit rate due to decay positrons reaches
1.4~MHz per sensor strip at the beginning of data taking near the muon
beam orbit, corresponding to a spatial density of
15~MHz/cm$^2$. We also developed front-end electronics with
a sampling rate of 200~MHz and a typical pulse width of less than
100~ns to withstand high hit-rate conditions~\cite{SliT128C}.
The number of hits from signal momentum tracks in the
range of 200--275~MeV/c is approximately 50.

\subsection{Performance evaluation of the detector module}

An experimental performance evaluation is essential to verify that the detector 
satisfies the required specifications. 
The basic requirements for operation in the J-PARC muon $g-2$/EDM
experiment have already been demonstrated using a prototype detector module~\cite{test_module}.
One of the remaining key performance 
metrics is the hit-rate capability under high hit-rate conditions, which is crucial for
properly reconstructing complete
positron tracks. The loss of signal hits due to pileup leads to failures in track reconstruction.
In this context, a muon beam test was carried out at the 
J-PARC MLF H-line using
the MuSEUM experimental setup~\cite{MuSEUM}.
The beam conditions available in the MuSEUM experiment are well suited for evaluating 
the hit-rate capability 
of the developed detector module under conditions relevant to the muon $g-2$/EDM experiment.
We demonstrate that the fabricated detector module satisfied
the required pileup tolerance under these conditions.

This paper presents the design of the developed detector module and 
an evaluation of its hit-rate capability based on 
muon beam data acquired in the MuSEUM experiment. Section~\ref{sec:detector} describes the
detector design and construction. Section~\ref{sec:experiment} summarizes the experimental setup 
and beam conditions during the MuSEUM beam test. The data analysis and performance evaluation
results are presented in Section~\ref{sec:performance}, followed by conclusions in Section~\ref{sec:conclusions}.

%% file: Detector.tex
\section{Detector system}
\label{sec:detector}
\begin{figure}[htbp]
    \centering
    \includegraphics[width=0.9\linewidth]{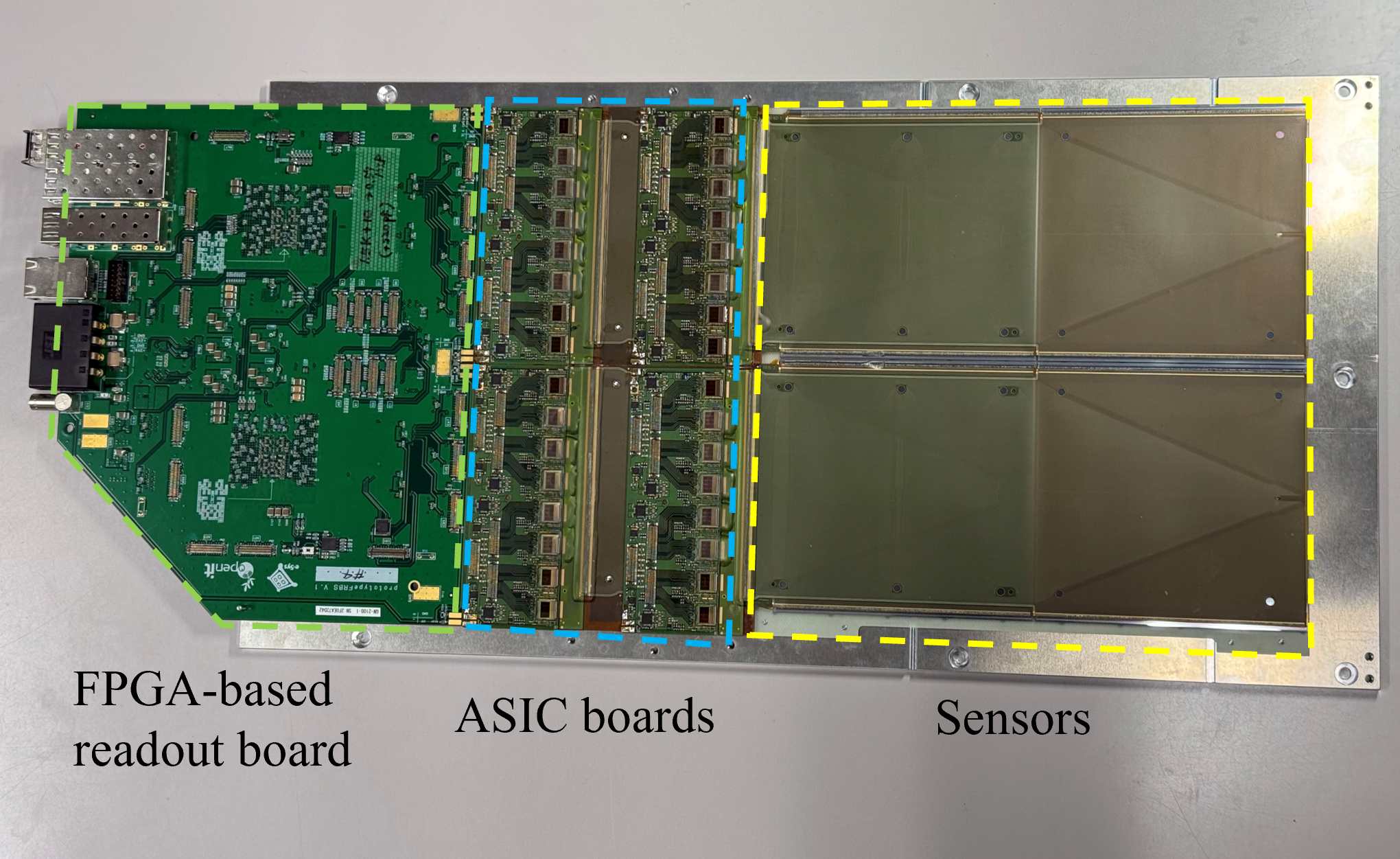}
    \caption{Produced quarter vane module. The quarter vane composed from four sensors,
    four ASIC boards and one FPGA-based readout board.}
    \label{fig:quarter_vane}
\end{figure}

A picture of the quarter vane is shown in Figure~\ref{fig:quarter_vane}.
The quarter vane is composed of sensors and
front-end electronics. The front-end electronics can be
divided into circuit boards containing readout ASICs (hereafter referred to
as ASIC boards) and
an FPGA-based readout board. Their features are described 
in the following subsections.

\subsection{Sensor}
\begin{figure}[htbp]
\centering
\includegraphics[width=0.5\textwidth]{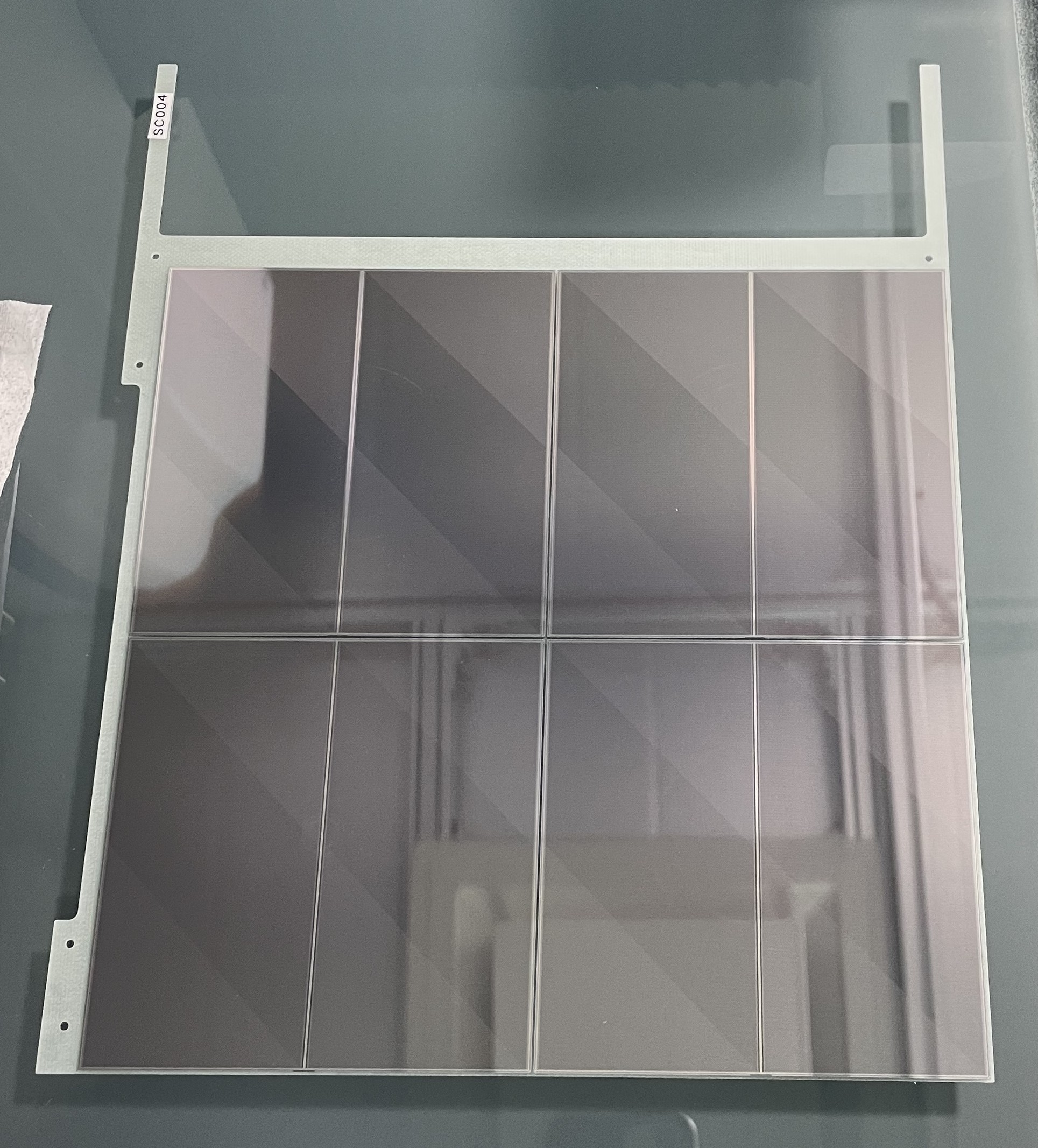}
\caption{Four silicon strip sensors glued on a glass-epoxy frame.}
\label{fig:sensor_component}
\end{figure}

A silicon strip sensor used in this detector module
is a single-sided p-on-n sensor with a double-metal
structure fabricated by Hamamatsu Photonics K.K. (model
S13804~\cite{S13804}). The sensor size is 98.77~mm
$\times$ 98.77~mm, with a thickness of 320~$\upmu$m.
The sensor has two columns of 512 strips, each with a 
pitch of 190~$\upmu$m
and a length of 48.365~mm. The full depletion voltage
is below 100~V, and the nominal bias voltage
is set to 120~V.
Four sensors are used in a quarter vane, as shown in
Figure~\ref{fig:sensor_component}, providing a detection area
of approximately 200~mm $\times$ 200~mm.
The performance of the sensor was evaluated using a prototype
detector module~\cite{test_module}. 
The detection efficiency was measured to be 99.8\%
at a nominal threshold of 0.3~MIP charge.
The time resolution of the sensor was measured to be 3.4~ns
at a nominal bias voltage of 120~V.

Signals from the sensor strips are routed via flexible printed circuit (FPC)
boards produced by Fujikura Ltd., which were specially designed
for this detector~\cite{Fujikura_FPC}.
The FPC boards are glued onto the sensor plane and connected to 
the readout ASICs described
in the next section.

\subsection{Front-end electronics}
\subsubsection{ASIC board}
To read out signals from the sensors, an ASIC called ``SliT'' 
has been developed.
Following the successful performance of the pre-production 
version, SliT128C~\cite{SliT128C},
the mass-production version, SliT128D, was fabricated.
Each ASIC chip has 128 readout channels. Each channel
includes an analog part consisting of a preamplifier, 
a CR-RC shaping
amplifier, and an additional CR circuit, which functions as a voltage
differentiator. The sensor strip is AC-coupled to the preamplifier,
and test pulses can be injected through a 100~fF coupling capacitor.
The pulse width of the CR-RC amplifier
is required to be less than 100~ns to withstand high hit-rate conditions,
and was measured to be 75~ns for a typical MIP charge.
The differentiator output provides a small time-walk signal
which is essential for the low-bias time measurement in 
the muon $g-2$ measurement.
Comparators are implemented at the outputs of both the CR-RC amplifier
and the differentiator, and their thresholds 
can be independently adjusted using 7-bit DACs.
The digital part consists of a signal interface, a memory controller,
a serializer, a timing generator, and a parameter controller.
The comparator outputs from the analog part are received,
and their timing is recorded using an external clock with
a nominal frequency of 200~MHz in a memory with a depth of 8192 
words per channel. Therefore, this corresponds nominally to a recordable 
time window of 40.96~$\upmu$s.
The falling edge of the comparator output is also recorded, enabling the determination of the
time-over-threshold (ToT) for each signal.

\begin{figure}[htbp]
\centering
\includegraphics[width=0.7\textwidth]{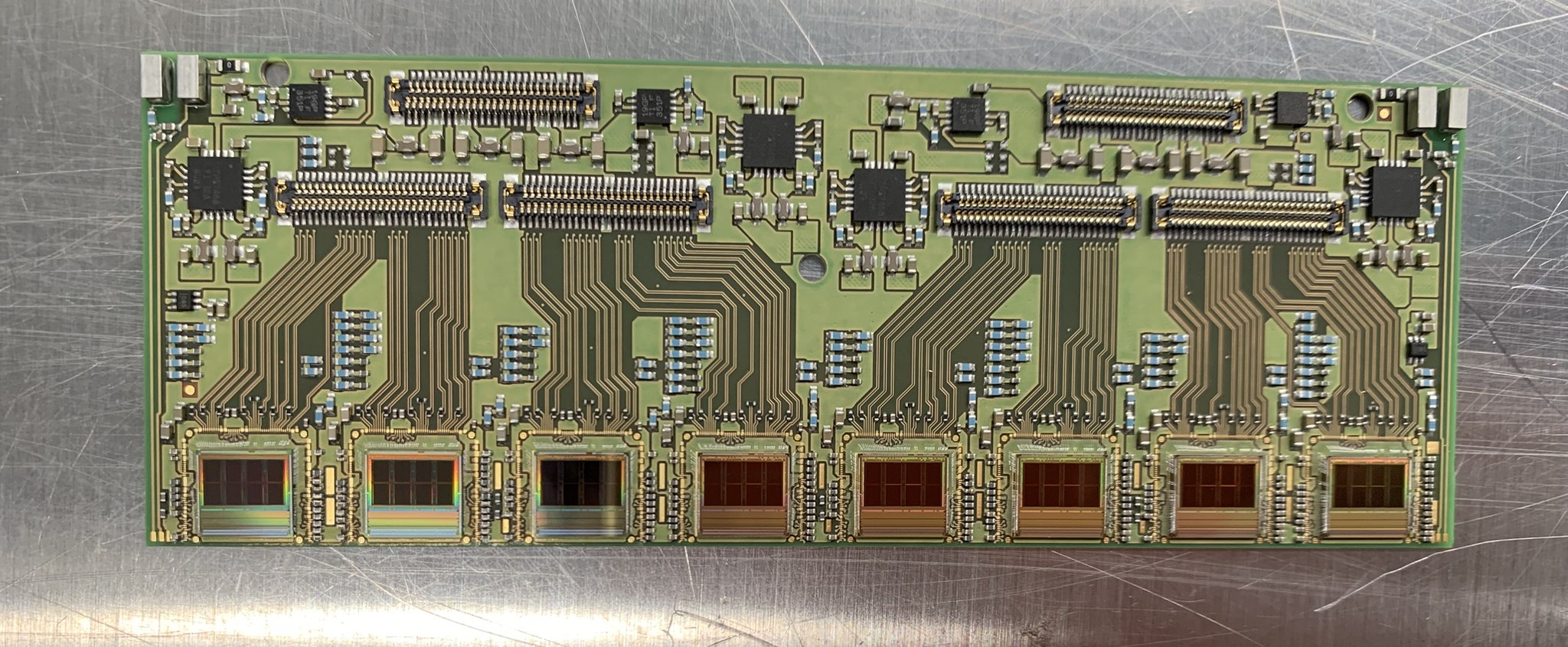}
\caption{Circuit board with ASICs. Eight ASICs are mounted on one board.}
\label{fig:ASIC_board}
\end{figure}

Eight SlIT128D chips are mounted on a single circuit board (see Figure~\ref{fig:ASIC_board}),
and four such boards
are used to read out all sensor strip signals in a quarter vane.
The ASIC boards are glued onto a copper plate, and heat pipes are 
attached to a copper cooling block
to dissipate heat. Cooling water is
circulated through the cooling block to maintain the operating
temperature.

\subsubsection{FPGA-based readout board}
The digital outputs of the SliT128D are processed by another circuit 
board, referred to as the
FPGA-based Readout Board for SliT (FRBS).
Two Xilinx Artix-7 FPGAs (XC7A200T-1FFG1156C) are used on a single FRBS to 
process the
outputs from 32 ASICs. The data are first stored in the
FIFO buffers implemented inside the FPGAs and are then  
serially transmitted to the backend electronics
via an SFP optical transceiver using the SiTCP protocol~\cite{SiTCP}.
The data from an FRBS is acquired by a system based on 
DAQ-Middleware~\cite{DAQ-Middleware}, which
is a software framework for network-distributed DAQ systems.

The ASIC boards and the FRBS are connected by micro-coaxial cables terminated
with XSLS-series connectors manufactured by KEL Corp.~\cite{KEL_XSLS}.
Heat pipes are attached to the metal packages of the FPGAs and are
thermally coupled to a copper cooling
block to dissipate heat.

%% file: Experiment.tex
\section{Experimental setup}
\label{sec:experiment}
\begin{figure}[htbp]
 \centering
 \includegraphics[width=0.8\textwidth]{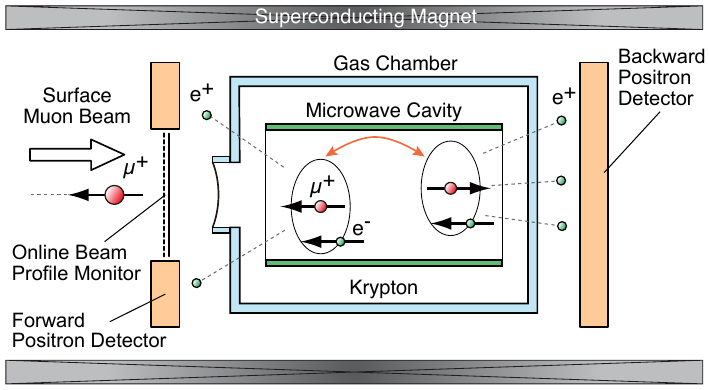}
 \caption{Schematic view of the MuSEUM experimental setup at high magnetic field. Reproduced from Ref.~\cite{MuSEUM}, \textcopyright The Author(s) 2025, licensed under CC BY 4.0.}
 \label{fig:MuSEUM_experiment}
\end{figure}

The MuSEUM experiment is conducted at J-PARC
to measure the ground-state hyperfine structure of muonium with high precision.
A schematic view of the experiment in a high magnetic field, which is carried
out at the J-PARC MLF H-line, is shown in Figure~\ref{fig:MuSEUM_experiment}.
An MRI-type solenoid magnet is installed along the beam line to
provide a magnetic field of 1.7~T, which splits the ground state of muonium via the
Zeeman effect. A gas chamber filled with krypton gas is placed inside the magnet. 
An almost 100\% polarized surface 
muon ($\upmu^{+}$) beam is injected into the gas chamber, and
when muons are stopped in the gas volume, muonium is formed through 
charge-exchange reactions with krypton atoms.
The muon spin can be flipped by applying a microwave magnetic field
into the microwave cavity located inside the gas chamber.
Muons decay with a lifetime of 2.2~$\upmu$s, and positrons from muon decays are emitted 
preferentially along the muon-spin direction. These positrons
are detected by detectors installed downstream and upstream of the gas chamber.
For precise measurement of the muonium hyperfine structure, high statistics data is necessary.
To achieve this, a high-intensity pulsed muon beam 
with an intensity of approximately $1\times 10^{8}~\upmu^{+}$/s available at the J-PARC H-line under 1~MW proton beam operation, is utilized~\cite{J-PARC_H-line}.
To cope with the large number of decay positrons from this intense muon beam, a detector 
with a high hit-rate capability is required.

The muon beam in J-PARC is
pulsed with a double bunch structure separated by 600~ns,
and the repetition frequency is 25~Hz.
In this study, one event was defined as a double-bunch muon beam pulse delivered at 25~Hz.
The beam test was carried out from late January
to mid-February in 2026. 
The proton beam power was approximately 710~kW in this period.
A quarter vane module was placed downstream of
the gas chamber. The module was wrapped in an
aluminum-foil noise-shielding tube firmly connected to electrical ground.
Between the gas chamber and the quarter vane, aluminum
plates with the total thickness of 55~mm were placed to absorb low momentum
positrons and electrons.

%% file: Performance.tex
\section{Performance evaluation}
\label{sec:performance}

Both the CR-RC amplifier output and the differentiator output were available 
on the quarter vane module, the former was used to evaluate the hit-rate capability.
The comparator threshold was set to 0.413~MIP charge based on the test pulse
calibration described below.
 
\subsection{Equivalent noise charge}
\begin{figure}[htbp]
    \centering
    \includegraphics[width=0.7\linewidth]{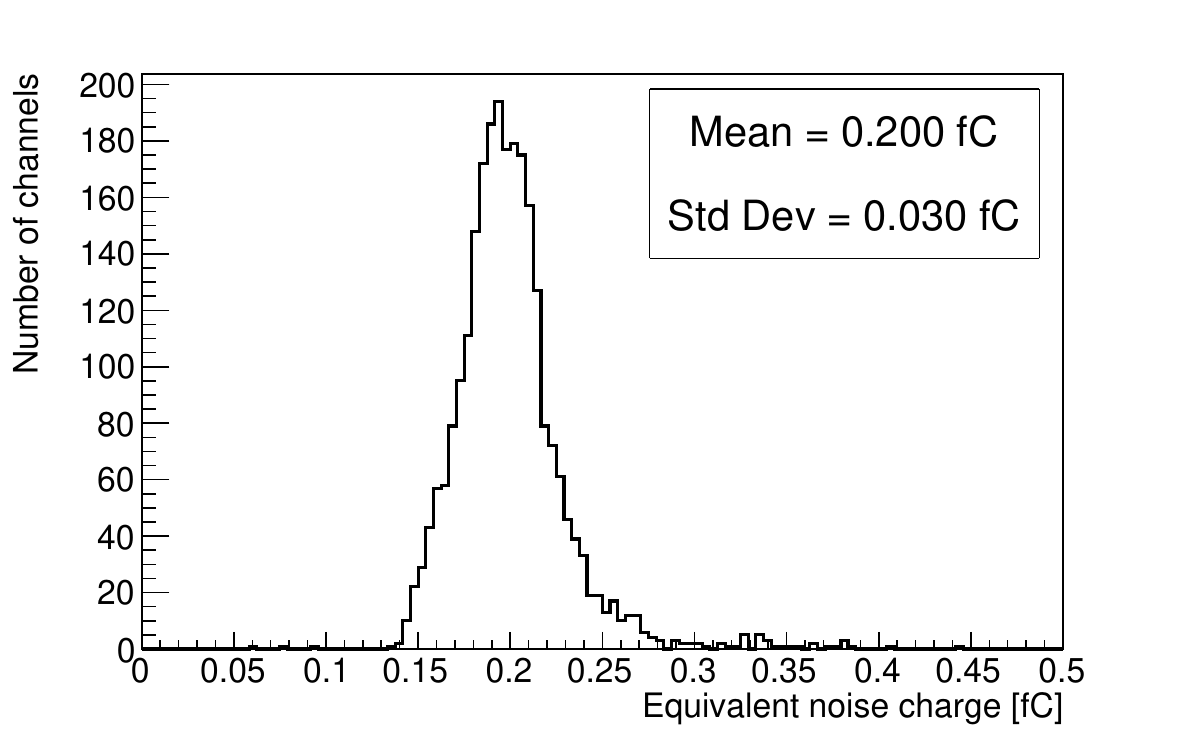}
    \caption{Equivalent noise charge distribution of ASIC channels on the quarter vane measured in the beam line after the shielding was installed.}
    \label{fig:noise_distribution}
\end{figure}

Prior to beam operation, the equivalent noise charge was 
measured to confirm that the electronics noise was
sufficiently small. The measurement was performed by injecting
a known test pulse charge into the circuit while scanning
the comparator threshold. The noise was estimated by fitting
an error function to the resulting efficiency-versus-threshold curve.
After the shielding was installed, the noise in the beam line
was reduced to 
the same level as that measured in the laboratory.
The measured equivalent noise charge distribution is shown in Figure~\ref{fig:noise_distribution}.
The average noise charge was 0.200~fC, which corresponds to approximately
1/18 of a MIP charge, indicating that the noise
contribution was sufficiently small compared to the signal.
However, it was found that the operation of an electrostatic kicker
in another beam line induced  
beam-synchronized noise in our detector. The following
evaluation was performed with this electrostatic kicker turned off.

\subsection{ToT-charge correlation calibration}
\begin{figure}[htbp]
    \centering
    \includegraphics[width=0.7\linewidth]{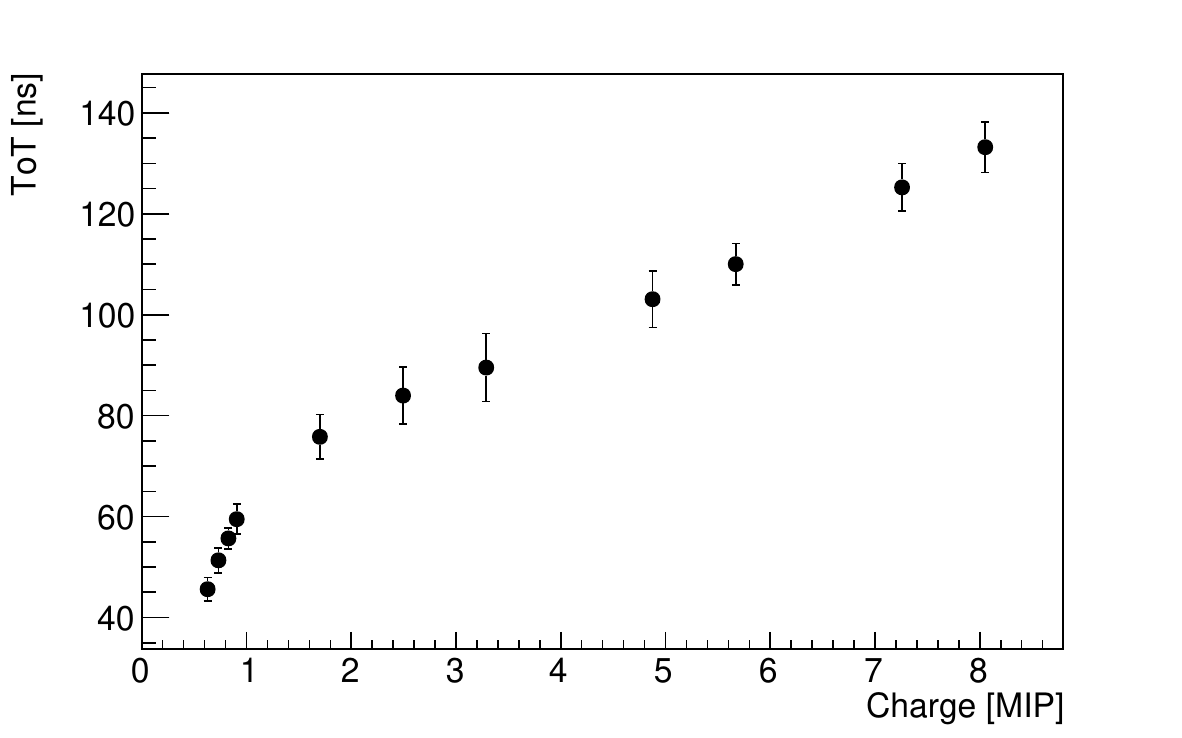}
    \caption{ToT averaged over all channels as a function of the injected charge,
    expressed in units of MIP charge. The error bars indicate the standard
    deviation across all channels.}
    \label{fig:ToT_vs_charge}
\end{figure}

The correlation between ToT and charge was calibrated by injecting test pulses with known amplitudes into the ASICs 
on the detector module, in order to determine the 
typical signal ToT response
for offline analysis.
The calibration was performed channel by channel. Figure~\ref{fig:ToT_vs_charge}
shows the ToT averaged over all channels as a function of injected charge.
The typical ToT for a single MIP charge was found to be 60~ns.

\subsection{Noise rejection}
\begin{figure}[htbp]
    \centering
\includegraphics[width=0.49\linewidth]{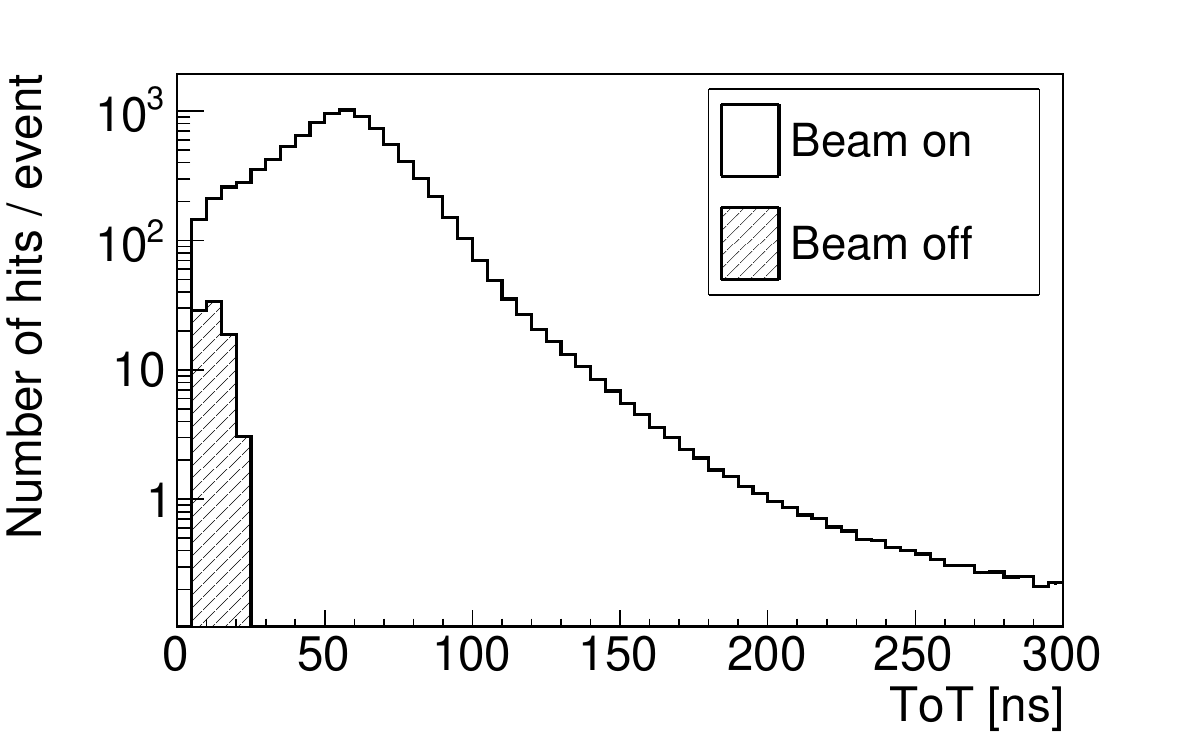}
\includegraphics[width=0.49\linewidth]{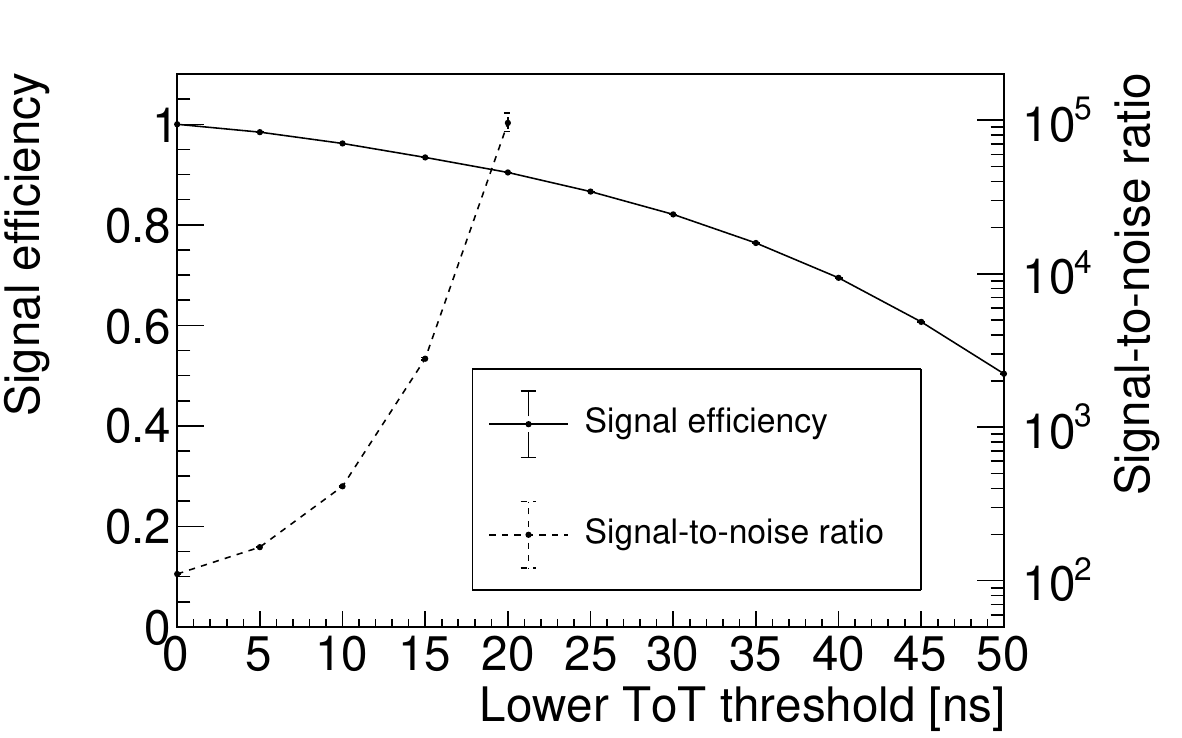}
    \caption{Distributions of the number of sensor hits as a function of ToT for beam-on and beam-off periods, normalized per event (left),
    and the signal efficiency
    and the signal-to-noise ratio as functions of the lower ToT threshold (right).
    For reference, the ToT corresponding to a single MIP is approximately 60~ns.}
    \label{fig:ToT_distribution}
\end{figure}

For offline analysis, it is important to 
evaluate and mitigate the contribution of non-beam-related noise.
Noise hits tend to appear at lower ToT values and can be estimated using data taken during beam-off periods.
Figure~\ref{fig:ToT_distribution} shows the distributions of the number of sensor hits as a function of ToT for beam-on and beam-off periods, normalized per event, together with the signal efficiency and the signal-to-noise ratio
as functions of the lower ToT threshold. To maximize the signal-to-noise ratio,
a cut of ToT~$>20$~ns was applied. After this ToT cut, the signal-to-noise 
ratio is greater 
than $9\times 10^{4}$ while maintaining a signal efficiency greater than 90\%.

\subsection{Hit rate capability}
To evaluate the hit-rate capability of the quarter vane module,
the distribution of the arrival times of decay positrons is analyzed. In the absence of a
resonant microwave field, the distribution is expected to follow an
exponential decay. If waveform pileup occurs in the analog readout
circuit, deviations from the exponential behavior are expected at early
times, where the hit rate is highest.

\begin{figure}[htbp]
\centering
\includegraphics[width=0.99\linewidth]{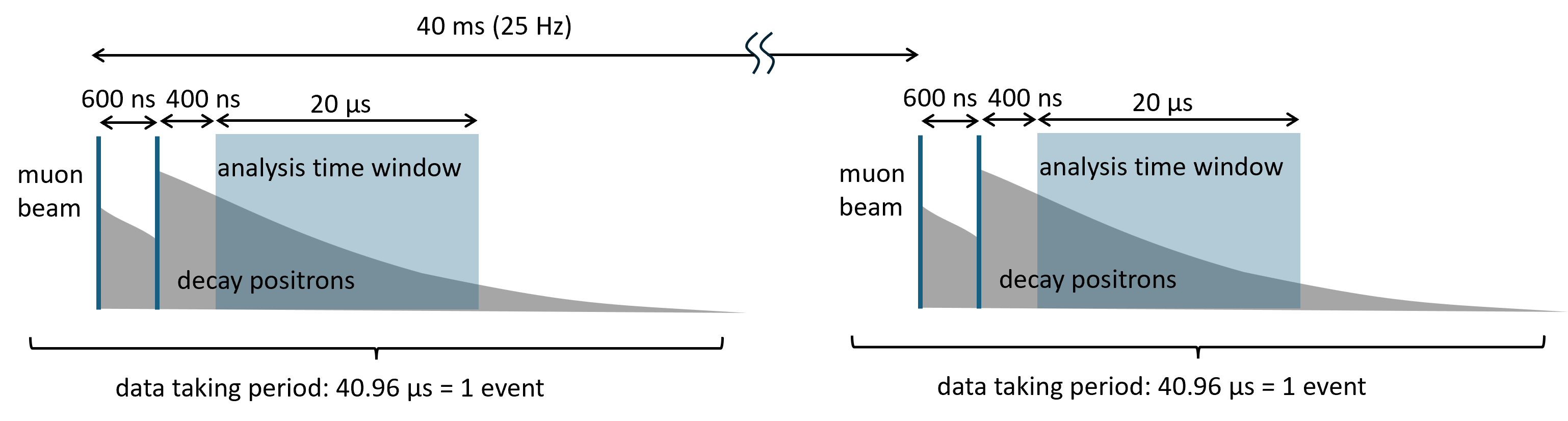}
\caption{Timing structure of the muon beam at the J-PARC MLF H-line and the measurement.}
\label{fig:timing_chart}
\end{figure}

The start of the analysis time window ($t=0$)
was set to 400~ns after the arrival of the second bunch of the muon beam, taking into
account the recovery time required after hits caused by positrons 
associated with the muon beam.
The length of the analysis time window was set to 20~$\upmu$s.
Timing structure of the measurement is shown in Figure~\ref{fig:timing_chart}.

\begin{figure}[htbp]
    \centering
    \includegraphics[width=0.7\linewidth]{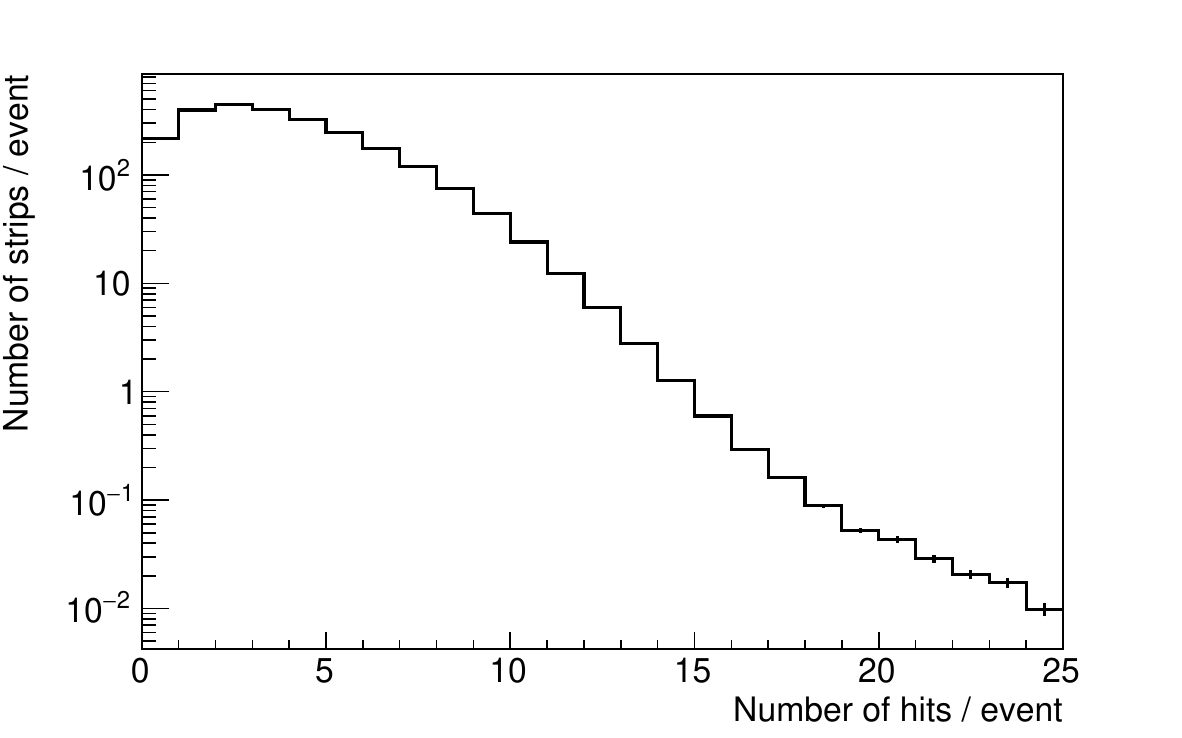}
    \caption{Distribution of the number of sensor strips as a function of the number of hits per event in a single strip within the analyzed
    time window, normalized per event.}
    \label{fig:nhits_dist}
\end{figure}

The data were categorized by the number of hits per event in a single strip 
within the analyzed time window.
The resulting distribution is shown in Figure~\ref{fig:nhits_dist}.
Assuming an exponential decay with the rest muon lifetime ($\tau$),
the number of hits per event ($N_{\mathrm{hits}}$) in a given time window ($T$) is expressed by
\begin{equation}
N_{\mathrm{hits}}=\int_{0}^{T}r_{0}e^{-\frac{t}{\tau}}{\mathrm{d}}t=r_{0}\tau\left(1-e^{-\frac{T}{\tau}}\right),
\end{equation}
where $r_{0}$ is the hit rate at $t=0$.
For $\tau=2.2$~$\upmu$s and $T=20$~$\upmu$s, the maximum hit rate is calculated
to be 0.455~MHz for 1 hit/event; thus,  
the targeted hit rate of 1.4~MHz corresponds to approximately 3~hits/event.

\begin{figure}[htbp]
    \centering
    \includegraphics[width=0.8\linewidth]{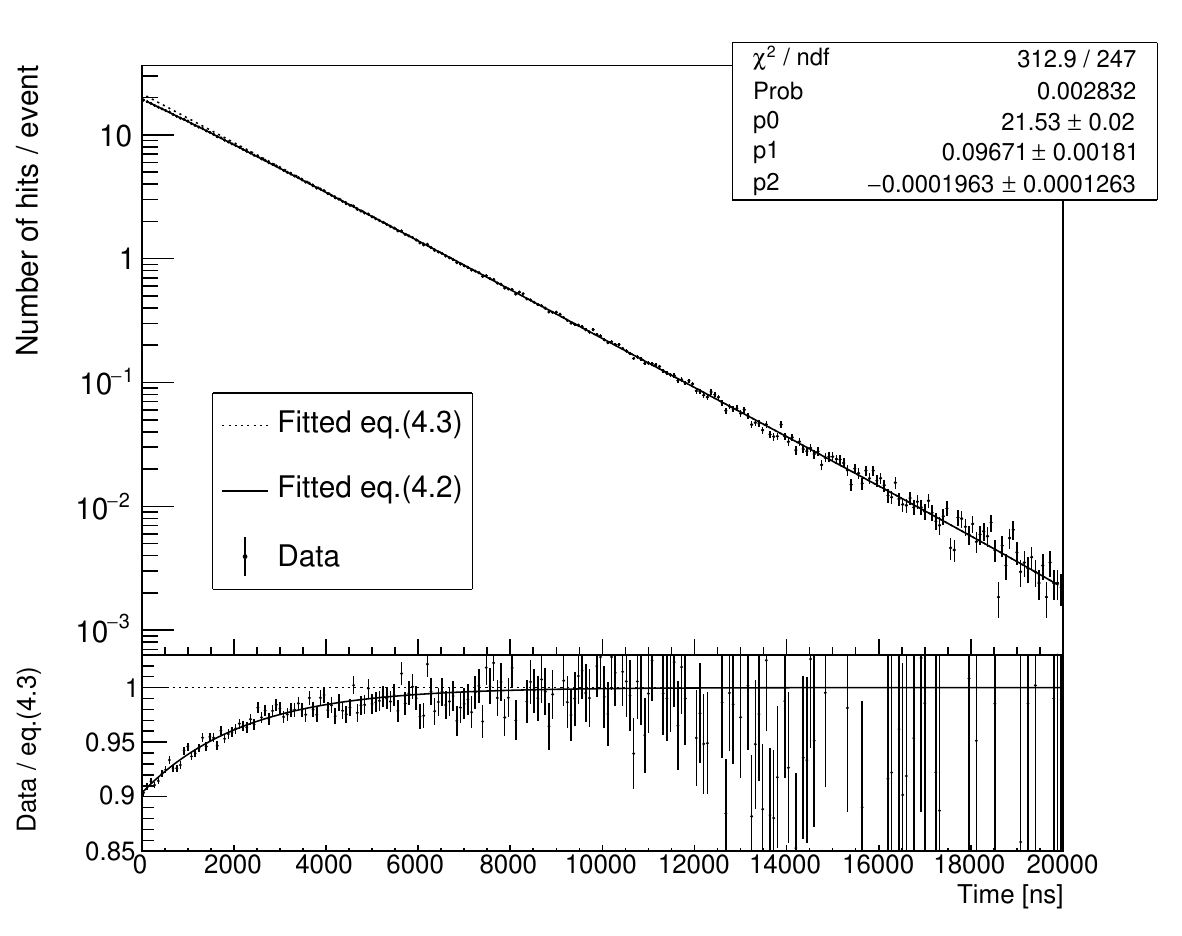}
    \caption{
    Distribution of the arrival times of 3-hit events within the 20~$\upmu$s time window,
    fitted with eq.~\eqref{eq:fit_function} using 80-ns bins.
    The ratio to eq.~\eqref{eq:exp_function} is shown at the bottom. The fit parameters are obtained from eq.~\eqref{eq:fit_function}.
    }
    \label{fig:time_spectrum_fit}
\end{figure}

For a given hit multiplicity, the observed arrival-time distribution of decay positrons was fitted with an exponential function multiplied by an efficiency-loss term due to pileup. If the pileup loss is proportional to the hit rate, the time distribution follows
\begin{equation}
    f(t)=p_{0}(1-p_{1}e^{-\frac{t}{\tau}})e^{-\frac{t}{\tau}}+p_{2}.
    \label{eq:fit_function}
\end{equation}
The parameter $p_{1}$ corresponds
to the maximum hit loss at $t=0$.
An example of a distribution fit is shown in Figure~\ref{fig:time_spectrum_fit}
for the 3~hits/event sample using eq.~(\ref{eq:fit_function}). In the fit,
$\tau$ is fixed to 2.1969811~$\upmu$s. To illustrate the pileup loss, the ratio
to the function without pileup loss,
\begin{equation}
    g(t)=p_{0}e^{-\frac{t}{\tau}}+p_2
    \label{eq:exp_function}
\end{equation}
is also shown at the bottom of the figure.
The obtained distribution is well described by eq.~\eqref{eq:fit_function}.

\begin{figure}
    \centering
    \includegraphics[width=0.8\linewidth]{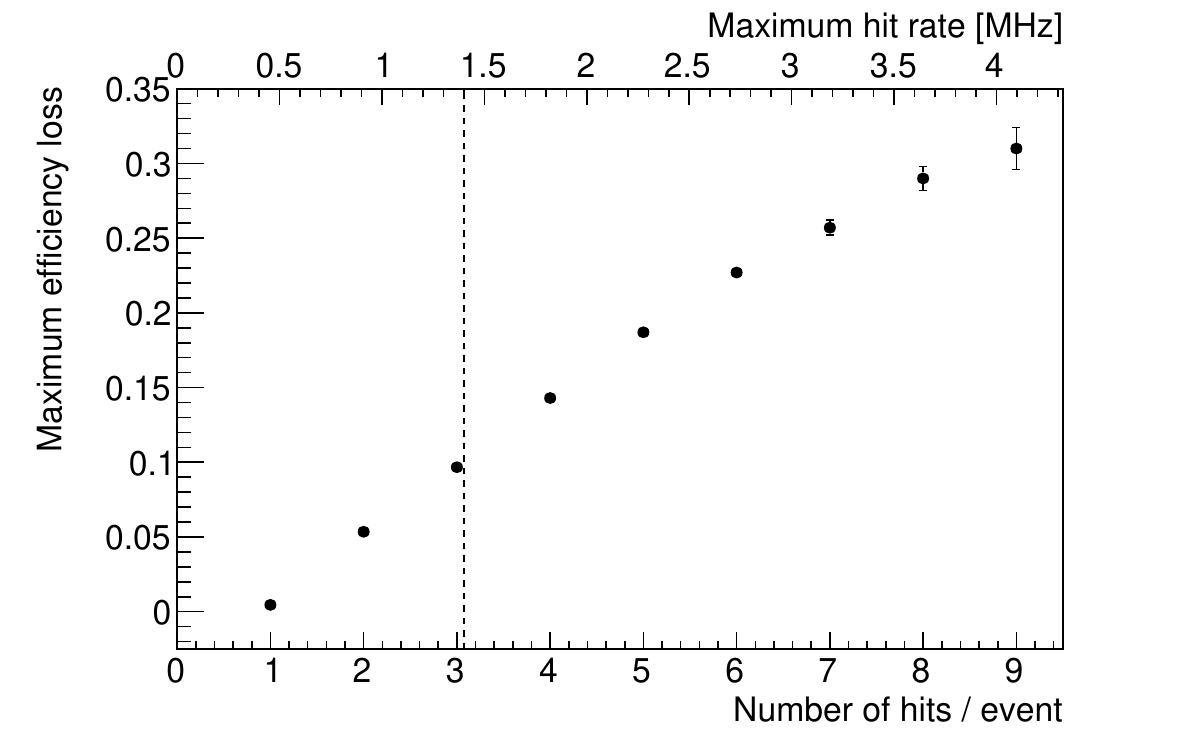}
    \caption{Maximum efficiency loss of hits as a function of the number of hits per event
    in a single strip within the analyzed time window.
    The corresponding maximum hit rate is also shown on the top axis.
    The highest expected hit rate of 1.4~MHz in the J-PARC muon $g-2$/EDM experiment is 
    indicated by a dashed line.}
    \label{fig:loss_vs_rate}
\end{figure}

Finally, the maximum efficiency loss (parameter $p_{1}$ in eq.~\eqref{eq:fit_function}) 
is plotted as a function of the number of hits per event in a single strip within
the analyzed time window, as shown in Figure~\ref{fig:loss_vs_rate}.
The figure also shows the maximum hit rate calculated from the number of hits per event.
At the expected hit rate of 1.4~MHz in the actual J-PARC muon $g-2$/EDM experiment,
the maximum efficiency loss is estimated to be 10\%. 
Considering that the expected number of hits associated
with signal tracks in the momentum range of 200--275~MeV/$c$ 
is approximately 50, 
a 10\% loss of hits is expected to have a negligible impact 
on track reconstruction performance.

The track reconstruction efficiency at the highest expected hit rate
has also been studied using simulations that include analog waveform modeling 
of the CR-RC amplifier~\cite{GPU_tracking}. A detailed consistency check between the simulation and 
measurement results is left for future work.

